\documentclass[aps,prd,groupedaddress,showpacs,nofootinbib]{revtex4}
\usepackage{graphicx,amssymb,bm}

\newcommand{\be}{\begin{equation}}
\newcommand{\ee}{\end{equation}}
\newcommand{\bea}{\begin{eqnarray}}
\newcommand{\eea}{\end{eqnarray}}
\newcommand{\beaa}{\begin{eqnarray*}}
\newcommand{\eeaa}{\end{eqnarray*}}
\newcommand{\nn}{\nonumber \\}

\begin{document}

\title{Ideal fluid and acceleration of the universe}

\author{
O. Gorbunova \footnote{Electronic address: gorbunovaog@yandex.ru}}

\address{Tomsk State Pedagogical University, 634041 Tomsk, Russia}

\begin{abstract}
The solution of the dark energy problem in models without scalars is presented.
It is shown that a late-time accelerating cosmology may be generated by an ideal fluid with
some implicit equation of state.
\end{abstract}

\maketitle

According to recent astrophysical data the current universe is expanding with
acceleration. Such accelerated behaviour of the universe is supposed to be due to the
presence of mysterious dark energy which ,at present, contributes about 70 procent of the
total universe energy-mass. What this dark energy is and where it came from is one
of the fundamental problems of modern theoretical cosmology
(for a recent review, see \cite{nabhan}, \cite{NojOd2}). Assuming a constant equation
of state (EOS), $p=w\rho$ , the dark energy may be
associated with a (so far unobserved) strange ideal fluid with negative
w. Astrophysical data indicate that w lies in a very narrow strip close to
 $w=-1$. The case $w=-1$ corresponds to the cosmological constant.
For $w$ less than $-1$ the phantom dark energy
is observed, and for $w$ more than $-1$ (but less than $-1/3$), the dark energy is described
by quintessence. It is interesting that the phantom phase is twice as probable than
the quintessence phase. Moreover, there are indications that there occured a recent
transition over cosmological constant barrier (over the phantom divide).

There are various approaches to describe the phantom dark energy. The simplest
one is to work with negative kinetic energy scalar (phantom).
The typical property of this phantom cosmology is a future singularity
which occurs in a finite time. This is due to the growth of phantom energy which may
lead to quite spectacular consequences. In particular, with growth of phantom energy,
typical energies (as well as curvature invariants) increase in the expanding universe. As
a result, in some scenarios the quantum gravity era may come back at the end of the
phantom universe evolution. In this case, it was checked that quantum effects
\cite{ElNojOd} ,\cite{NojOd6},\cite{Srist}, \cite{NojOd8} may act against the Big Rip
(or even stop it in case of quantum gravity \cite{ElNojOd}).
The interesting approach to describe the dark energy universe is related to an
ideal fluid with some (strange but explicit) EOS which may be sufficiently complicated
\cite{NojOd9}, \cite{NojOd10}, \cite{CapCord}.
Of course, this approach is phenomenological in some sense, because it does not
describe the fundamental origin of dark energy. At the same time, it may lead to quite
successful description of not only phantom phase but also of transition from decelation to
acceleration or crossing of the phantom divide.

We now consider the FRW cosmology with an ideal fluid. The starting FRW universe
metric is:

\be ds^2=-dt^2 + a(t)^2
%\sum_{i=1}^3 \left(dx^i\right)^2\
.\label{FRW} \ee

In the FRW universe, the energy conservation law can be expressed as \be
\label{ppH1} 0=\dot\rho + 3H\left(p + \rho\right)\ . \ee Here
$\rho$ is the energy density, and  $p$ is the pressure. The Hubble rate  $H$
is defined by $H\equiv \dot a/a$ and the first FRW equation is
\be \label{pH3} \frac{3}{\kappa^2}H^2=\rho\ . \ee
We often consider the case that  $\rho$ and $p$ satisfy the simple EOS, $p=w\rho$.
Then if  $w$ is a constant, Eq. (\ref{ppH1}) can be easily integrated as
 $\rho=\rho_0 a^{-3(1+w)}$.
Using the first FRW
equation
(\ref{pH3}), the well-known solution follows: \be \label{ppH4} a=a_0 \left(t -
t_1\right)^\frac{2}{3(w+1)}\quad \mbox{or} \quad a_0 \left(t_2 -
t\right)^\frac{2}{3(w+1)}\ , \ee when $w\neq -1$, and \be
\label{ppH5} a=a_0\,e^{\kappa t\sqrt{\frac{\rho_0}{3}}} \ee when
$w=-1$.

The ideal fluid with a more general EOS may be considered
\cite{NojOd10}, \cite{CapCord}: \be \label{EoS1} f(\rho, p)=0\ . \ee
An interesting example is given by
\be \label{EoS2} \rho\left(1 + \frac{A}{2}(\rho +
p)\right) + \frac{3}{\rho}\left(2 + \frac{A}{2}(\rho + p)\right)^2
=0\ .\ee Solving Eqs.(\ref{ppH1}) and (\ref{pH3}) one arrives at \be
\label{EoS3}  H=\frac{1}{t}\left( 1 -
\frac{\kappa^2t^2}{A}\right)\ ,\quad \rho =
\frac{3}{\kappa^2t^2}\left( 1 - \frac{\kappa^2t^2}{A}\right)^2\ ,\nn
 p=\frac{1}{\kappa^2}\left(\frac{1}{t^2} - \frac{8\kappa^2}{A} +
\frac{3\kappa^4 t^2}{A^4}\right)\ . \ee Since \be \label{EoS4}
\dot H= - \frac{1}{t^2} - \frac{\kappa^2}{A}\ , \ee it follows that
\be \label{EoS5} \frac{\ddot a}{a}=\dot H + H^2 =
\frac{\kappa^2}{A}\left(- 3 + \frac{\kappa^2 t^2}{A}\right)\ .\ee
Then if  $A>0$, the decelerating universe with $\ddot a<0$
transits to an accelerating universe
with
 $\ddot a>0$ when
$t=\sqrt{\frac{3A}{\kappa^2}}$. Eq.(\ref{EoS4}) also shows that if
 $A<0$ - the non-phantom phase with $\dot H<0$
changes to the phantom phase with $\dot H>0$ and
$t=\sqrt{-\frac{A}{\kappa^2}}$.

This demonstrates that
an ideal fluid with a complicated EOS may be the origin of dark energy and late-time
acceleration.

\ 

\noindent
{\bf Acknowledgements} 
This work is done in collaboration with Profs. S. Nojiri and S.D. Odintsov.
The research is supported by LRSS project N4489.2006.02

\end{document}